\documentclass[twocolumn,showpacs,amsmath,amssymb]{revtex4}

\usepackage{graphicx}
\usepackage{dcolumn}
\usepackage{bm}
\usepackage{amssymb}
\usepackage{latexsym}
\usepackage{amsfonts}

\newcommand{\IV}{$I-V $ }
\begin{document}

\title{Structural and electrical characterization of hybrid metal-polypyrrole nanowires}

\author{L. Gence}
\altaffiliation{Corresponding author.}
\affiliation{Institut de la Mati\` ere Condens\' ee et des Nanosciences - Bio \& Soft Matter (IMCN/BSMA), Universit\' e catholique de Louvain, Place Croix du Sud, 1, B-1348 Louvain-La-Neuve, Belgium}

\author{V. Callegari}
\affiliation{Institut de la Mati\` ere Condens\' ee et des Nanosciences - Bio \& Soft Matter (IMCN/BSMA), Universit\' e catholique de Louvain, Place Croix du Sud, 1, B-1348 Louvain-La-Neuve, Belgium}

\author{O. Reynes}
\affiliation{Laboratoire de Chimie Inorganique, Universit\'e Paul Sabatier, 118 Route de Narbonne, 31062 Toulouse,  France}

\author{S. Faniel}
\affiliation{Information and Communication Technologies,
Electronics and Applied Mathematics - (ICTEAM/ELEN), Universit\' e catholique de Louvain, Place Croix du Sud, 1, B-1348 Louvain-La-Neuve, Belgium}

\author{S. Melinte}
\altaffiliation{Research associate of the FNRS.}
\affiliation{Information and Communication Technologies,
Electronics and Applied Mathematics - (ICTEAM/ELEN), Universit\' e catholique de Louvain, Place Croix du Sud, 1, B-1348 Louvain-La-Neuve, Belgium}

\author{C. Gustin}
\affiliation{Agence de Stimulation Technologique, Rue du Vertbois, 13b
B-4000 Li\`ege }

\author{V. Bayot}
\affiliation{Institut de la Mati\` ere Condens\' ee et des Nanosciences - Nanophysics (IMCN/NAPS), Universit\' e catholique de Louvain, Place Croix du Sud, 1, B-1348 Louvain-La-Neuve, Belgium}

\author{S. Demoustier-Champagne}
\altaffiliation{Research associate of the FNRS.}
\affiliation{Institut de la Mati\` ere Condens\' ee et des Nanosciences - Bio \& Soft Matter (IMCN/BSMA), Universit\' e catholique de Louvain, Place Croix du Sud, 1, B-1348 Louvain-La-Neuve, Belgium}

\date{\today}


\begin{abstract}
We present here the synthesis and structural characterization of hybrid Au-polypyrrole-Au and Pt-polypyrrole-Au nanowires together with a study of their electrical properties from room-temperature down to very low temperature. A careful characterization of the metal-polymer interfaces by transmission electron microscopy revealed that the structure and mechanical strength of bottom and upper interfaces are very different.
Variable temperature electrical transport measurements were performed on both multiple nanowires - contained within the polycarbonate template - and single nanowires.  Our data show that the three-dimensional Mott variable-range-hopping model provides a complete framework for the understanding of transport in PPy nanowires, including non-linear current-voltage characteristics and magnetotransport at low temperatures.
\end{abstract}
\pacs{72.20.Ee, 72.20.Ht, 73.40.Ns, 73.63.-b}
\keywords{}
\maketitle
\section{Introduction}
\label{sec:Introduction}
In the recent years, an increasing number of studies have centered on nanowires,
mainly due to their potential use in nanoelectronics~\cite{Review}.
It has been demonstrated that  nanowires can be used in various applications such as molecular sensors~\cite{Cui01a}, field effect transistors~\cite{Cui01b, HJChung05} or in nano-electro-mechanical systems~\cite{Husain03}.
The template method is a common and versatile technique for producing nanowires and it is probably the
simplest way to synthesize rod-shaped multi-segmented nanostructures~\cite{Koezuka83, Nishizava95}.
One  interesting aspect of this method is that it permits the growth of in-wire organic junctions by sequential deposition of the metallic and the organic layers~\cite{Mbindyo02}. Among the organic materials studied for their electronic properties in one-dimensional environments, conjugated polymers have received  special attention~\cite{HJChung05, Park04, Long05, Aleshin04, Park02b, He03, Saha02}. 
Different transport mechanisms have already been proposed in  conjugated polymer nanostructures, depending on the synthesis conditions and geometry of the devices. Despite these  experimental efforts, the electrical transport  at  nanoscale in conjugated polymers is still matter of debate.
\begin{figure}[ht!]
\centering
\includegraphics[width=0.9\columnwidth]{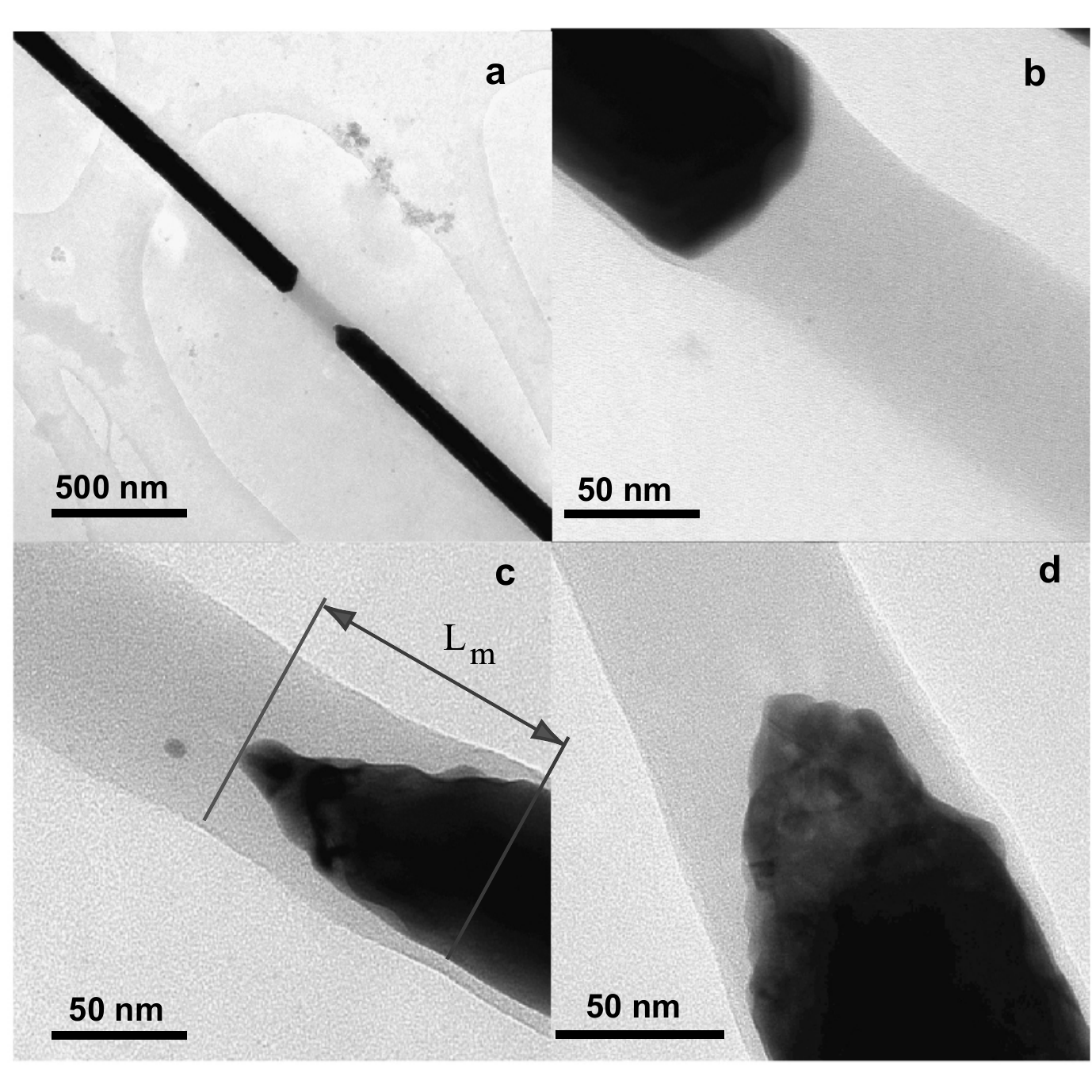}
\caption{TEM characterization of Au-PPy-Au nanowire interfaces. (a) Typical TEM picture of a 70 nm diameter Au-PPy-Au nanowire. Panel (b) displays its bottom (first grown) interface, while panel (c) presents its upper (second grown) interface. The size of the gold meniscus $\rm L_m$  is indicated. (d) Upper interface of a 90 nm nanowire.}
\label{Fig1}
\end{figure}
Until now, very few studies~\cite{HJChung05, Park04} focused on the synthesis and electrical characterization of hybrid conjugated polymer-metal  nanowires. 
Moreover, none of them reported on the structural characterization of the metal-polymer interfaces, which could play an important role on the electrical properties of the nanodevices.  
We present  here the synthesis and  structural characterization of hybrid metal(Au or Pt)-polypyrrole(PPy)-metal(Au) nanowires
together with a study of their electrical properties from room-temperature down to very low temperature ($T \approx 0.5 \  \rm K$).  
Metallic segments, few microns long, serving as  contacting electrodes are electrochemically synthetized within polycarbonate (PC) templates, before and after the PPy nanowire growth. Interestingly, the resulting metal-PPy-metal hybrid nanowires have two morphologically different PPy-metal interfaces (see Fig. 1). In terms of mechanical robustness the PPy-onto-metal interface is smoother and more fragile than the metal-onto-PPy interface.  By using Pt  instead of Au for the first grown segment, much more robust nanowires have been fabricated.
The present all-electrochemical technique allows controlled optimization of the mechanical strength of the PPy-metal interfaces, produces high-quality electrical contacts, and permits a thorough electrical characterization of nanowires in both vertical and planar geometry.
Variable temperature  electrical measurements of multiple polymer nanowires embedded in the PC template are compared to those of single nanowires. Similar sets of data have been obtained. In all samples, current-voltage \IV characteristics are symmetrical and show no rectification effect.
They are linear from room temperature down to $T \approx 100 \  \rm K$ and  non-linear at lower temperatures.
Our data show that the three-dimensional Mott variable-range-hopping (VRH) model provides a complete framework for the understanding of transport in PPy nanowires. Two interesting facets of this conduction mechanism have been investigated in more details: the electric field-induced hopping and positive magnetoresistance.
\begin{figure}
\centering
\includegraphics[width=0.9\columnwidth]{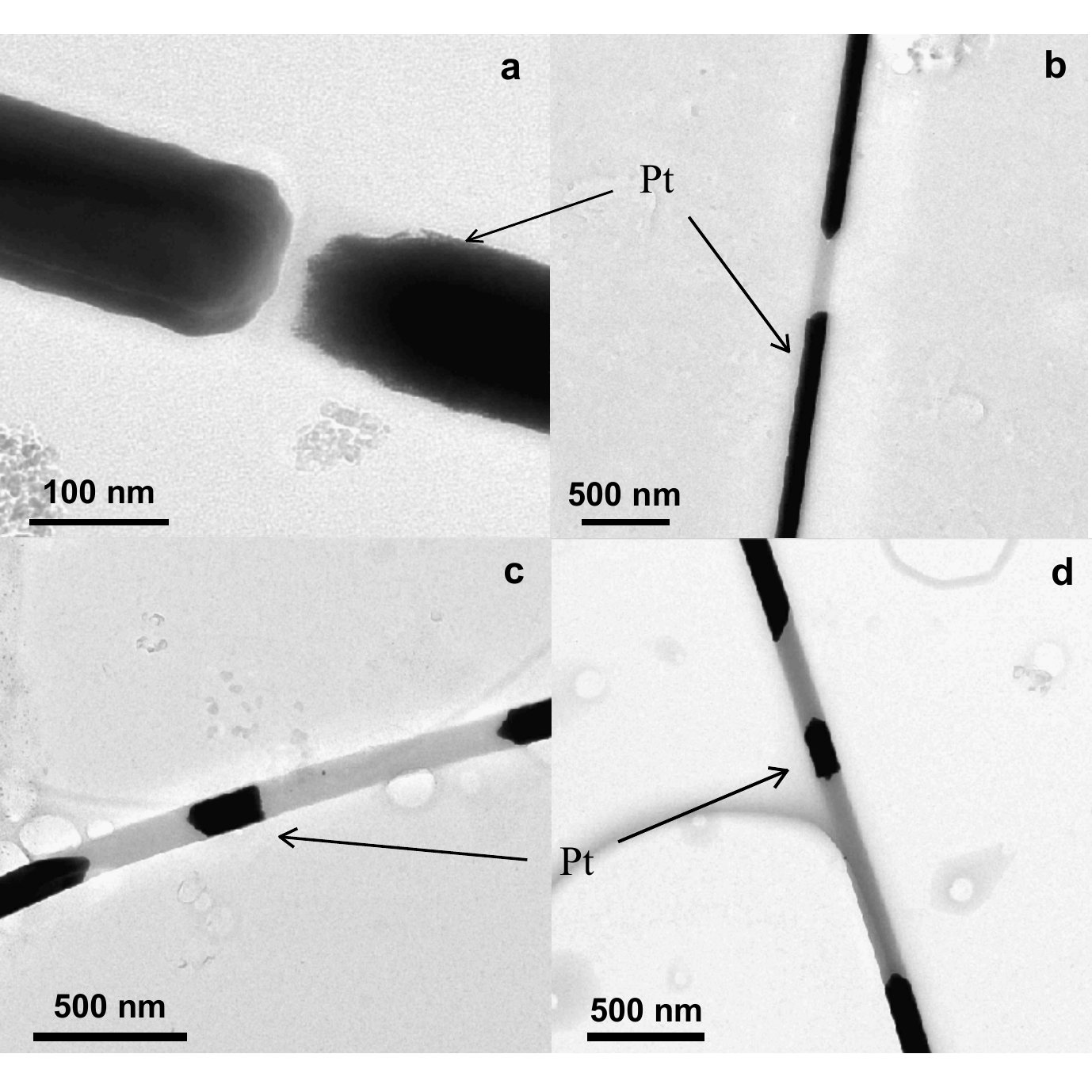}
\caption{(a and b) TEM pictures of tri-segmented Pt-PPy-Au nanowires displaying  very short PPy in-wire junctions. (c and d) TEM pictures of multi-segmented Au-PPy-Pt-PPy-Au nanowires.}
\label{Fig2}
\end{figure}
\section{Experimental details}
\label{sec:Experiment}
Segmented composite Au (or Pt)-PPy-Au nanowires presenting various diameters and PPy lengths were efficiently prepared by an all-electrochemical template method described in detail elsewhere~\cite{Reynes05,Dauginet04}.
Briefly, the electrochemical growth was carried out  in an one-compartment cell at room temperature with a Pt disc counter electrode and an Ag/AgCl reference electrode.
Nanoporous PC track-etched membranes with  a thickness of $21 \rm \ \mu m$, pore diameters  ($\phi$'s) of 70 and 90 nm, and a pore density of $10^{9} \rm /cm^{2}$ were supplied by it4ip~\cite{it4ip} and used as templates. An approximatively 300 nm-thick  layer of Au evaporated on one side of the membrane served as working electrode.
Electropolymerization of pyrrole was carried out by cyclic voltammetry in de-ionized water in presence of 0.1M $\rm LiClO_{4}$, 0.01M pyrrole, and $7 \times 10^{-4}$M sodium dodecylsulfate used as surfactant to enhance the membrane wetting. To get PPy nanowires (and not nanotubes) the potential was swept between 0 and 0.85 V at a scan rate of 300 and $400 \rm \ mV.s^{-1}$ for nanostructures with diametres of 70 and 90 nm, respectively. The length of the PPy segment can be easily adjusted at the desired length by controlling the number of potential scans. Typically, one hundred scans lead to a PPy segment length of 600 nm. Electrodeposition of Au was performed by cycling the potential of the working electrode from 0.7 to 0~V at $200 \rm \ mV.s^{-1}$ using a home-made free-cyanide solution (0.1M KCl, 0.1M $\rm K_{2}HPO_{4}$, and 0.03M of $\rm HAuCl_{4}$ in de-ionized water). Generally, 100 scans lead to an Au segment with length of $\approx 5\  \mu \rm m$.
The Pt segment was electrodeposited potentiostatically at  $-0.2$  V and the  plating solution was made in-house from 0.01M $\rm Na_{2}PtCl_{6}.6H_20$ and 0.5M $\rm H_{2}SO_{4}$ in de-ionised water.
The resulting nanowires are well aligned within the membrane and their surface is smooth compared to nanowires produced with commercial membranes~\cite{Duvail}.

The  morphology of the hybrid nanowires has been studied by scanning electron microscopy (SEM) and by transmission electron microscopy (TEM) after dissolution of the PC template. The separation of the nanowires is as follows. After the growth, the membranes were soaked several times in methanol and de-ionized water. Then, the membranes were dissolved by immersion in a dichloromethane solution containing dodecyl sulfate and the suspension was placed in an ultrasonic bath to separate the nanostructures from the Au film evaporated on one side of the membrane. Occasionally, the suspensions were filtered through poly-(ethylene terephthalate) membranes with pore diameters ranging between 0.8 and $\approx 3 \  \mu \rm m$. In order to remove any contaminant from the nanowire surface, the samples were thoroughly rinsed with dicholoromethane. The specimens for TEM analysis were  prepared by placing a drop of the nanowire suspension on a carbon grid. Scanning and transmission electron  images were obtained using a High Resolution FEG Digital Scanning Microscope Gemini 982  and a Transmission Microscope Gemini 922, respectively.
X-ray photoelectron spectroscopy (XPS) data were obtained using a SSI X probe (SSX 100/206) spectrometer from Fisons, equipped with and Al anode (10~keV) and a quartz monochromator. 

We have electrically characterized both multi-nanowire samples embedded in the PC membrane and single nanowire devices. The measurements were performed in a two terminal configuration by applying a DC voltage to the sample and measuring the current with an electrometer. Multi-nanowire devices were investigated from 300 to 0.5 K, while single nanowires were studied from 300  to 4 K. For temperatures below 4 K, the total dissipated power was kept below 100 nW corresponding to a power of approximately 1 pW per nanowire. Taking a typical value of the low-$T$ thermal conductivity for amorphous polymers~\cite{Choy77}, we estimate that the sample's heating is negligible down to 0.5 K.
Measurements of nanowires embedded in PC membranes were performed by contacting both sides of the membrane with silver paste. Empty templates have a resistance above $10^{12} \ \Omega$ and templates filled with gold nanowires have a typical resistance lower than $50 \ \Omega$.
The planar single nanowire devices were fabricated as follows. As previously described, after the dissolution of the PC membrane, the nanowires were dispersed in dichloromethane. This solution was dropped on $1 \ \rm \mu m$-wide electrical contacts consisting of 5 nm of Ti as an adhesion layer and 35 nm of Au, defined by photolithography on  $\rm SiO_{2}$ layers thicker than 30 nm obtained via dry oxidation. Leak currents lower than $1$ pA were measured for empty  planar devices over the voltage range scanned in the experiments~\cite{Substrates}. 
Optical microscopy was used to identify the contacted nanowires instead of electron microscopy to avoid PPy radiation damage. Corresponding contact pads were connected to the sample holder pins with silver paint and $20 \  \mu \rm m$ gold wires.  
Test devices consisting of electrodeposited single gold nanowires  attached to Au contact pads were used to estimate the contact resistances.
The room-temperature total resistance (including leads and contact resistance) measured for single $3 \ \rm \mu m$ long Au nanowires with diameter of  70 nm is less than $600 \ \Omega$.  Using a standard bulk value of resistivity for Au, it can be deduced that the contact resistance between the Au segment of the nanowire and the Au contact pads is less than $500 \ \Omega$, which is negligible compared to the resistance of one PPy nanowire ($\approx 10^{8} \ \Omega$).

\begin{table}
\begin{tabular}{c|ccc|ccc}
\hline
$\phi$ (nm) & &70 &   && 90 &\\
\hline
PPy length (nm)                                  & 670      & 1680        & 3110         & 915        & 1440    & 3340  \\
$\rm L_m$ (nm)                        & 66         & 75            & 88              & 85          & 180       & 225 \\
Ad. surface ($\rm nm^{2}$)    & 9680    & 11000     & 12900       & 16400   & 33740  & 42600 \\
\hline
\hline
\end{tabular}
\caption{Length of the PPy segment for various Au-PPy-Au nanowires. For these nanowires, the gold meniscus length ($\rm L_m$) and the corresponding calculated adhesion surface of the upper interface are reported.}
\label{Tab1}
\end{table}

\section{The polymer-metal interface}
\label{sec:Interface}

Tri-segmented Au-PPy-Au or Pt-PPy-Au nanowires with PPy junction lengths from 20 nm to 3 $\mu \rm m$ were fabricated by sequentially electrodepositing the metal and the polymer within the nanopores of PC templates under electroplating conditions described in the experimental section. The length of each nanowire segment (Au, Pt, and PPy) is monitored by adjusting the amount of electrical charge passed during the electrochemical process. The resulting hybrid nanostructures can be freed from the template membrane and collected as an ensemble of free particles for observation by electron microscopies.  Typical TEM pictures of these hybrid nanowires are shown in Figs. 1 and 2. These pictures unambiguously demonstrate the absence of short-circuits, even with very short PPy segments (down to 20 nm). However, the observation that upon dispersion onto patterned substrates around 50\% of hybrid  Au-PPy-Au nanowires were missing one metallic section, lead us to investigate the strength and the morphology of both PPy-metal interfaces in our devices.

Observations by TEM, summarized in Fig.~\ref{Fig1}, reveal that the two interfaces of the hybrid Au-PPy-Au nanowires have quite different shapes.
The bottom interface is relatively flat, indicating that the first metal section growth is homogeneous under our synthesis conditions. On the other hand, for the upper  interface, a conical gold meniscus soars into the PPy section, ensuring a strong mechanical link between the polymer and the upper metal section. This  results in a larger adhesion surface compared to that of the bottom interface,
explaining why the metal-onto-polymer interface is mechanically more robust than the polymer-onto-metal interface.
The presence of this meniscus could be explained by pore wall effects (essentially electrostatic interactions between the growing polycation chains and the anionic sites along the membrane pore walls), similar to those leading to the formation of PPy nanotubes~\cite{Reynes05,Dauginet04}.
In the present work, the natural tendency of PPy to grow along the pore walls was thwarted  by properly adjusting  pyrrole electropolymerization rate.  The creation of the meniscus 
seems to occur as soon as the electropolymerization is initiated and it has been observed in all investigated samples displaying various PPy junction length. 
The gold meniscus length $\rm L_m$ has been measured by TEM for several  upper interfaces of 70 and 90~nm diameter Au-PPy-Au nanowires with  different PPy junction lengths. The results are  reported in Table~\ref{Tab1}.  The corresponding adhesion surfaces are also given in Table~\ref{Tab1}.
For both nanowire sizes, $\rm L_m$ and the adhesion surface increase with the length of PPy junction. Nevertheless, $\rm L_m$ and  the adhesion surface are smaller for all the 70 nm diameter nanowires compared to the  90 nm diameter nanowires. The adhesion surface roughly correlates with the percentage of unbroken nanowires dispersed onto devices and it is thus an important parameter in assessing the mechanical robustness of the nanowires.
 
Interestingly, by changing the nature of the first metallic segment, it was possible to grow stronger bottom polymer-onto-metal interfaces.  Indeed, tri-segmented Pt-PPy-Au nanowires were prepared by an analogous all-electrochemical procedure. By optical inspection, upon dispersion of the as-prepared nanowires, we can estimate that the percentage of broken nanowires is less than 20\%.  Thanks to that significant improvement of the bottom  interface strength, Pt-PPy-Au segmented nanowires with very short polymer in-wire junction (as small as 20 nm) were successfully prepared [see Fig.~\ref{Fig2}(a)]. Although not in the focus of the present paper, we illustrate in Fig.~2(c and d) that hybrid nanowires with complex structures could be grown due to the controlled synthesis of Pt, Au, and PPy segments.
\begin{figure}
\centering
\includegraphics[width=1\columnwidth]{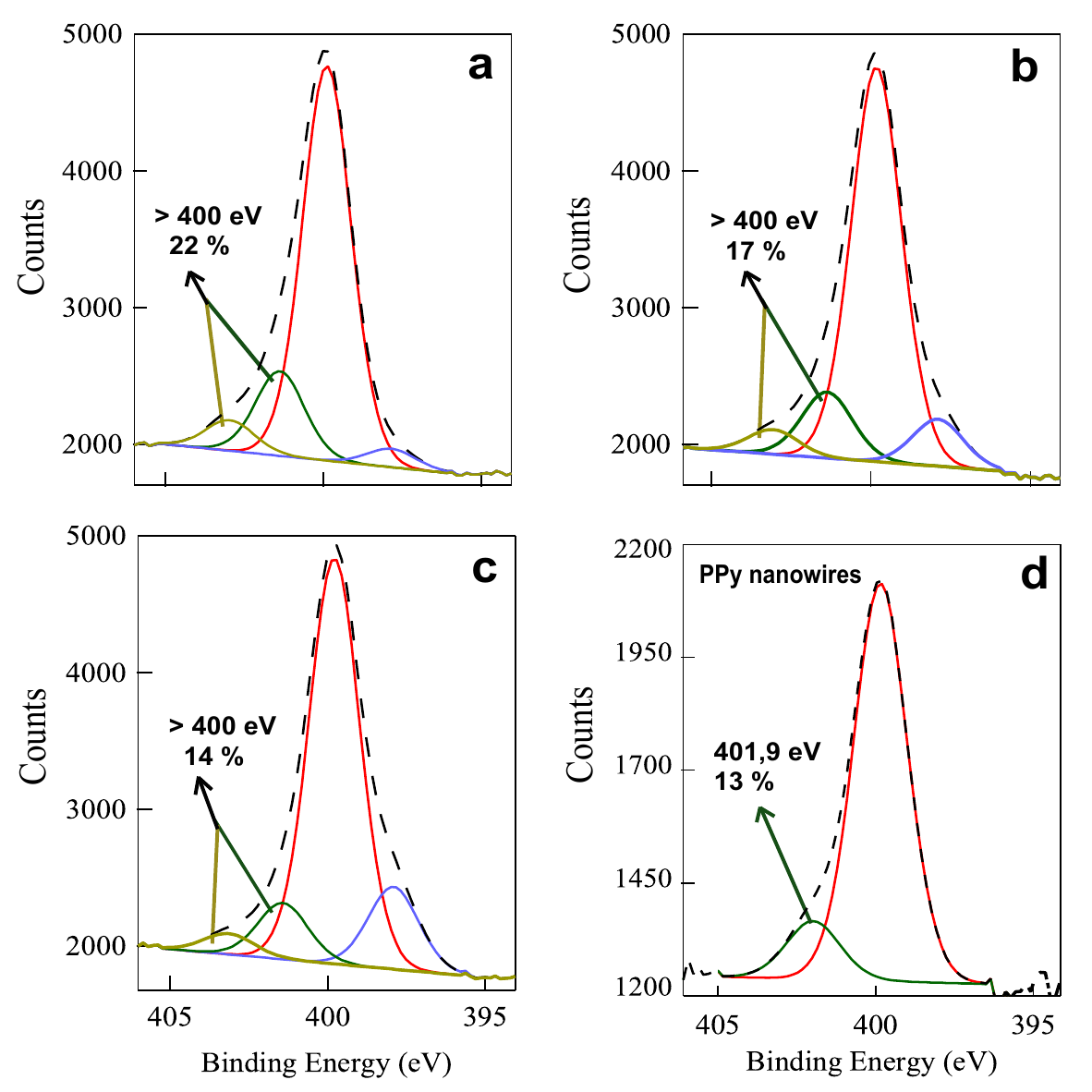}
\caption{[Color online] N(1s) XPS fitted spectrum of PPy films: as-synthesized (a), after application of a reduction potetial of -0.2 V for 10 min. (b), after application of a reduction potetial of -0.2 V for 30 min. (c) and N(1s) XPS fitted spectrum of one-component PPy nanowires (d).}
\label{Fig3}
\end{figure}

Besides the structure of the metal-polymer interfaces, the PPy doping ratio is another key parameter playing on the electrical transport properties of the hybrid nanowires. XPS is a particularly useful tool for gathering information on the doping level of electroactive polymers. In particular, previous XPS studies on PPy demonstrated that the imine like (=N-), amine like (-NH-), and positively charged nitrogen atoms ($\rm -N^{+}$) corresponding to any particular intrinsic oxidation state and protonation level of the polymer can be quantitatively differentiated in the  N(1s)  XPS spectrum~\cite{Kang93, Dauginet05}. A typical N(1s) XPS fitted  spectrum of a PPy film, electrosynthesized under the conditions described in the experimental section, is shown in Fig. 3(a).  Two components (green curves) - representing protonated nitrogens - appear at binding energies higher than 400 eV, while two other components, corresponding to amine-like (red curve) and imine-like (blue curve) non-protonated nitrogens, appear at 399.8 and 397.9 eV, respectively. The doping level in the as-synthesized PPy film, estimated by the [N+]/[N] ratio, is close to 22\%. Then, XPS measurements were performed for one-component PPy nanowires electrosynthesized under similar conditions [Fig. 3(d)].  In this case, the N(1s) curve could be simply fitted by only two components~\cite{fit}. The first peak (green curve) representing protonated nitrogens is centered at 401.9 eV and the second one (red curve) corresponding to non-protonated nitrogens  appears at 399.8 eV. The doping level for one-component PPy nanowires was estimated at 13\% (one protonated nitrogen out of eight). As already reported for conjugated polymer nanotubes [18,20], the lower doping value for template-grown PPy nanostructures compared to PPy films results from the electrosynthesis in nanopores. For tri-segmented metal-PPy-metal nanowires, one expects a {\it lower} value of the doping ratio ($< 13\%$) due to the reduction  of the PPy segment during the electroplating of the upper metal  segment. To flesh out  the scale of this reducing process,  XPS has been used to estimate the doping in PPy films after application of a post-treatment reproducing the conditions  applied for the electrodeposition of the Pt segment onto the PPy segment~\cite{tri}.  Specifically, a potential of -0.2 V was applied to PPy films during 10 or 30 minutes.  As shown in Fig.~3(b and c), a progressive decrease of the protonated nitrogen components (green curves) combined to a marked increase of the non-protonated imine-like nitrogen (blue curve) is observed in the XPS data.  This corresponds to a decrease of the doping ratio from 22\% to 17\% and further to 14\% after application of reduction conditions for 10 and 30 minutes, respectively.  Based on these results, we can estimate that for tri-segmented metal-PPy-metal nanowires  the doping ratio of PPy is only few percents, in agreement with the electrical behavior described in the following section. 

\begin{figure}
\centering
\includegraphics[width=0.9\columnwidth]{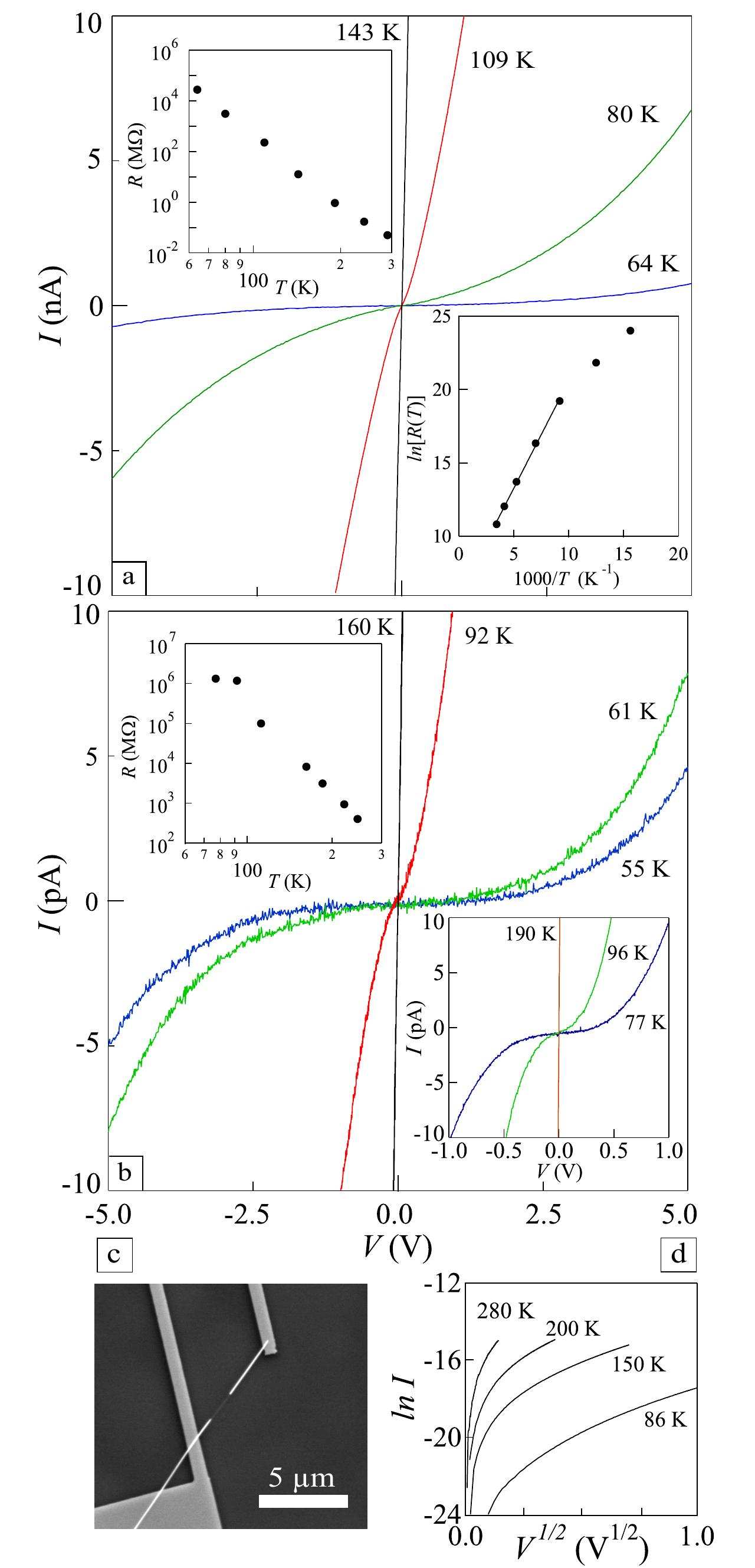}
\caption{[Color online] (a) $I-V$ characteristics for the sample M90-1. Upper inset shows the $T$ dependence of the resistance. Lower inset represents  $ln [R(T)]$ versus $1/T$. As illustrated, $E_a$ could be deduced from a linear fit to the high-$T$ data points.
(b) $I-V$ characteristics of sample S70. Upper inset shows the $T$-dependence of the resistance. Lower inset gives the \IV characteristics of  sample M70-2. (c) SEM picture of a nanowire placed on two gold leads. (d) $ln \  I $ versus $V^{1/2}$ plots at various $T$'s for sample M90-2.}
\label{Fig4}
\end{figure}

\section{Electrical transport}
\label{sec:Electrical}
We present here electrical measurements for six specimens: S70 (single nanowire, $\phi = 70\ \rm nm$), M70-1 and M70-2 (membranes, $\phi = 70\ \rm nm$), M90-1, M90-2, and M90-Pt (membranes, $\phi = 90\ \rm nm$). While samples S70, M70-1, M70-2, and M90-1 have a nominal PPy length of $1500 \rm \ nm$, the PPy segment is reduced to $300 \rm \ nm$ for samples M90-2 and M90-Pt. Only sample M90-Pt contains Pt-PPy-Au nanowires. Sample M70-2 includes a self-assembled thiol monolayer onto the first grown Au segment. In Subsection IV.A  the \IV characteristics are displayed as a function of $T$ and the transport mechanisms are discussed. We further analyze our data in Subsection IV.B, where we present our magnetotransport measurements. In Subsection IV.C we analyze the electric field-induced hopping regime and in Subsection IV.C we discuss the relantionship between the electrical measurements and the nanowires' structure.

\subsection{\IV characteristics}
The  current-voltage \IV spectroscopy is a powerful technique to gain insight into various transport phenomena such as tunnelling and rectification.
Typical current-voltage \IV characteristics  are given in Fig.~\ref{Fig4} for various temperatures  $T \gtrsim 50\ \rm  K$ and different specimens.
In all samples, the \IV plots are symmetrical and show no rectification effect. 
The \IV characteristics are ohmic between room temperature and approximatively $T \approx  \rm 120 \ K$ [see $T= 143 \ \rm K$ in Fig. 4(a) - sample M90-1,  $T= 160 \ \rm K$ in Fig. 4(b) - sample S70, and $T= 190 \ \rm K$ in lower inset to Fig. 4(b) - sample M70-2].
Noteworthy, in Pt-PPy-Au samples  the ohmic region extends down to liquid nitrogen temperatures.
As shown in Fig.~\ref{Fig4}, the non-linearity of the \IV curves increases with decreasing temperature, signaling that a peculiar conduction mechanism occurs at very low temperatures (see  below).
The resistance values $R(T)$ obtained from the $(dI/dV)^{-1}$ values at zero bias are shown in the upper insets to Fig.~\ref{Fig4}(a and b) for samples M90-1 and S70, respectively. 
The resistance has been calculated only for $T \gtrsim 50 \ K$. A zero-current plateau develops  in the \IV characteristics at lower $T$, precluding the determination of samples' resistance. 
For all investigated samples, the resistance monotonically increases with decreasing $T$, indicating that PPy has a non-metallic behavior.
The room-temperature conductivity of $\approx 0.04 \ \rm S.cm^{-1}$ obtained for single Au-PPy-Au nanowires is comparable to that of  bulk insulating polypyrrole. This could be understood in the light of our XPS results which indicate a low doping level of our PPy samples.
\begin{table*}
\begin{center}
\begin{tabular}{c|cccccc}
\hline
 &   $  \qquad T_{0}$  ($\rm K)\qquad $&$N(E_{F}) (\rm eV^{-1}.cm^{-3}) \qquad $& $\alpha^{-1} (nm)\qquad$& $F_{0}$ ($\rm V.m^{-1}) \qquad $& $\alpha^{-1}(F_0,T_0)$ (nm)&  \\
\hline
S70  &1.4x$10^{8}$ &-&-  &1.5x$10^{13}$&2.1\\
M70-1  &2.4x$10^{8}$  & - &- &-&-\\
M90-1 &1.8x$10^{8}$ &-&-  & 1.8x$10^{13}$&0.9\\
M90-2  &6.2x$10^{7}$ & 1.1x$10^{18}$&1.4 & 1.5x$10^{13}$&1.0\\
M90-Pt &7.6x$10^{6}$ & 2.6x$10^{18}$&2.1  & 0.1x$10^{13}$&1.8\\
\hline
\hline
\end{tabular}
\caption{Data analysis for the  samples S70, M70-1, M90-1, M90-2, and M90-Pt. 
The  Mott temperature $T_{0}$ was obtained by fitting the $ln [R(T)]$ versus $T^{-1/4}$ plots and $F_{0}$ was extracted from linear fits to $ln \ I \propto (F_0/F)^{1/4}$ data (see text). The parameter $N(E_{F})$ is estimated within the 3D VRH model and the parameter $\alpha^{-1}(F_0,T_0)$ is estimated within the 3D field-induced hopping model.}
\label{Tab2}
\end{center}
\end{table*}

At low voltage and high temperature, current is usually attributed to thermally excited carriers. This mechanism yields an ohmic conduction exponentially dependent on temperature~\cite{Sze81}. 
Hence, we have  plotted $ln [R(T)]$ versus $1/T$ for $T \gtrsim 50 \  \rm K$ and extracted the activation energy $E_a$. The deduced activation energies are $E_a$ = 121, 133, 130, 92, and 52 meV for samples S70, M70-1, M90-1, M90-2, and M90-Pt, respectively. These  values compare well with the values previously reported~\cite{Park04}. However, the activated transport cannot explain our low-$T$ data, as shown in the lower inset of Fig.~\ref{Fig4}(a),  as well as the non-linear characteristics appearing at higher excitation voltage and lower temperature.\\

In hybrid metal-PPy-metal nanowires, electron transport could be dominated by carrier injection through the metal-PPy interface.
The injection-dominated mechanisms are $T$-independent tunneling and $T$-dependent thermionic emission. Schottky plots consisting of $ln [I/T^{2}]$ versus $1/T$ at various fixed $V$'s and $ln \ I$ versus $V^{1/2}$  at different temperatures were used to check if the conduction is electrode limited. Our data can not be casted within the model of Schottky emission. As illustrated in Fig.~\ref{Fig4}(d) for sample M90-2,  $ln \ I$ does not follow a linear relationship with $V^{1/2}$.  This is not surprising since it is expected that PPy forms ohmic contacts with metals with a high work function like Au and Pt~\cite{Koezuka83}. 
\begin{figure}
\centering
\includegraphics[width=0.9\columnwidth]{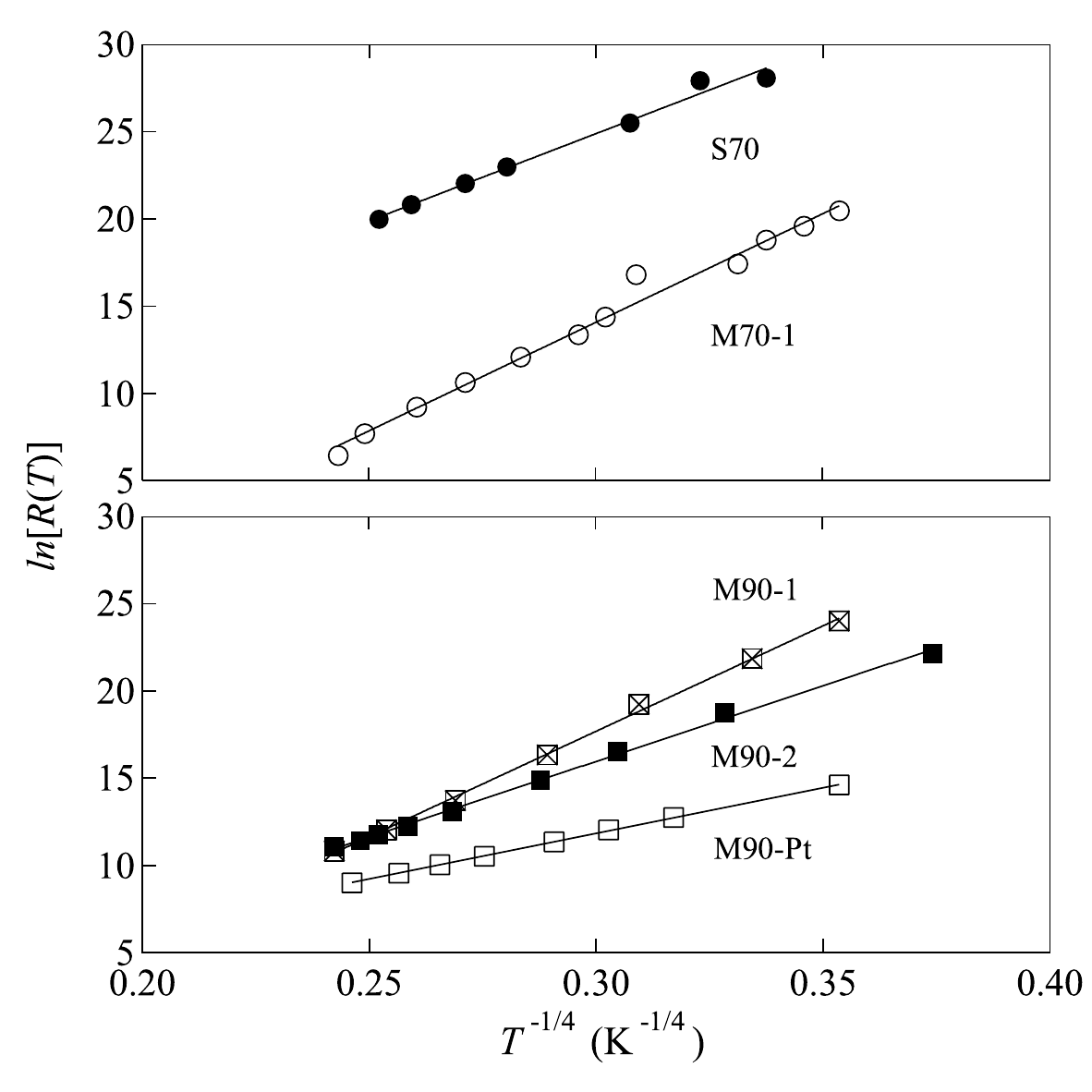}
\caption{Three-dimensional VRH plots (see text). The solid lines are fits to the data.}
\label{Fig5}
\end{figure}

Recently, several groups have studied the low $T$ transport in various polymer nanowires such as polyacetylene or PPy~\cite{Aleshin04,Shen06}.  The transport properties have been analysed in terms of different models such as Luttinger liquid or quasi-1D variable range hopping.  These models refer to one-dimensional tunneling processes that cannot satisfactorily explain the $T$-dependence of the resistance in our PPy samples.   Besides the different nature of the polymers under scrutiny in these studies, the fact that electrical transport could be cast within one-dimensional models has various origins such as the effective dimensionality of the system, the amount of disorder, the level of doping, and the configuration of the electrical probes. These models are not directly applicable for our weakly doped and strongly disordered nanowires, eventhough their diameter is below 100 nm.\\

We now show that transport in our samples, can be explained by a thermally activated tunneling among states that are localized in a constant density of states~\cite{Mott79}. The VRH mechanism is a general form of charge transport for insulators and usually dominates over the thermally-activated transport at sufficiently low $T$. In the VRH model, the resistance follows the relation $ln [R(T)] \propto (T_{0}/T)^{1/(d+1)}$, where $T_{0}$ is the Mott temperature and  $d$ is the dimensionality of the system. The best fits to our data have been obtained with $d=3$. This suggests that the three-dimensional (3D) VRH is the appropriate model of transport in our samples.  The $ln [R(T)]$ versus $T^{-1/4}$ plots are given in Fig.~\ref{Fig5} for $T \gtrsim 50 \  \rm K$. From these plots we extract the parameter $T_0$. The results are summarized in Table~\ref{Tab2}. In the case of 3D VRH,  $T_{0} = 16\alpha^3/k_{B}N(E_{F})$. Here $k_{B}$ is the Boltzmann constant, $\alpha^{-1}$ is the localization length, and $N(E_{F})$ is the density of states at the Fermi energy. 
The localization length is linked to the amount of disorder in PPy and determines, together with $N(E_{F})$, the Mott temperature $T_{0}$. In the following Subsection B, we show that $\alpha^{-1}$ can be deduced from low-$T$ magnetoresistance data (Fig.~\ref{Fig6}).
Typical low temperature  \IV characteristics  are shown in Fig.~\ref{Fig7} for the S70 sample.
Clearly, as $T$ is lowered, a zero-current plateau develops in the  \IV curves and the current becomes temperature independent ($T \lesssim 10 \rm \ K$). This behavior can be related to the electric field-induced hopping conduction (see Subsection C), which follows the non-ohmic hopping conduction mechanism as $T$ decreases. 

\subsection{Magnetoresistance measurements}
Commonly,  when the VRH mechanism governs the electrical conduction, the presence of the magnetic field $B$ results in a positive magnetoresistance $R(B)$.
For the low-$B$ range, the wavefunction shrinkage model developed by Shklovskii and Efros \cite{Shklovskii84} predicts that $ln \left[R(B)/R(0)\right]=t(B/B_{c})^{2}\left(T_{0}/T\right)^{1/4}$, where $t$ is a numerical constant ($t \approx 0.1$) and $B_{c} =6\hbar/[e\alpha^{-2}(T_{0}/T)^{1/4}]$ is the critical magnetic field.
In the high-$B$ limit, a $B^{1/3}$ dependence of $ln \left[R(B)/R(0)\right]$ is expected.
This high magnetic field range corresponds to $B \gg B_{c}$.
\begin{figure}
\centering
\includegraphics[width=0.9\columnwidth]{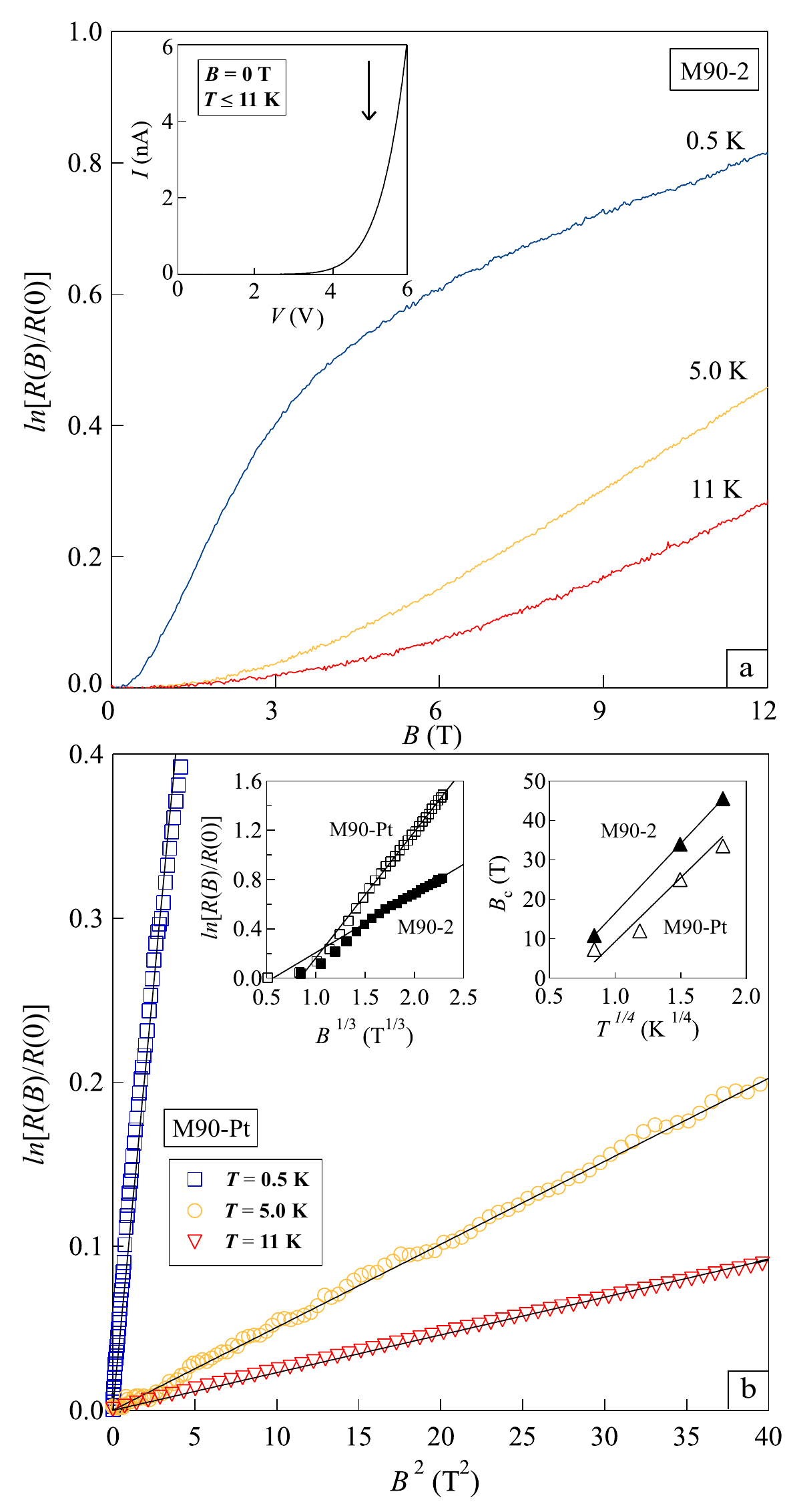}
\caption{[Color online] (a) Variation of $ln{[R(B)/R(0)]}$ with  $B$ at indicated $T$'s for the sample M90-2. Inset: Below $T = 11 \ \rm K$ the \IV curves are indistinguishable. (b) Variation of $ln{[R(B)/R(0)]}$ as a function of $B^{2}$ in the low magnetic field range at $T$ = 11, 5, and 0.5~K for the sample M90-Pt.  The solid lines are fits to the data. The left inset shows $ln{[R(B)/R(0)]}$ as a function of  $B^{1/3}$  for samples M90-2 and M90-Pt at $T = 0.5 \ \rm K$. The lines are guides to the eye. The right inset displays $B_{c}(T)$ as a function of $T^{1/4}$ for both samples (see text). The solid lines are fits to the data.}
\label{Fig6}
\end{figure}

We present magnetoresistance measurements for samples M90-2 and M90-Pt performed with $B$ aligned along the nanowires axis~\cite{MR}.
The plot of $ln \left[R(B)/R(0)\right]$ with respect to $B$ for sample M90-2 is given in Fig.~\ref{Fig6}(a) at $T =11$, 5, and 0.5 K.
The magnetoresistance is measured close to the $I=0$ plateau of the $I-V$ curves [see arrow in inset to Fig.~\ref{Fig6}(a)].
Several features are noteworthy in our data: (i) the observed magnetoresistance is always positive, (ii) it is symmetric in magnetic field: $R(B,T)=R(-B,T)$, (iii) it increases as $T$ is lowered, (iv) and it displays a parabolic-like behavior at low $B$ for all investigated temperatures.

These observations suggest that our magnetoresistance data could be cast within the wavefunction shrinkage model. 
The left inset to Fig.~\ref{Fig6}(b) shows that, at high $B$, the magnetoresistance follows the expected $B^{1/3}$ dependence.
The low-$B$ data is displayed in Fig.~\ref{Fig6}(b) as  $ln{[R(B)/R(0)]}$ versus $B^{2}$  at indicated temperatures.
The $B^{2}$ dependence is observed up to few Tesla. From the linear fits to these data, we  extract $B_{c}(T)$, which is plotted as a function of $T^{1/4}$ in the right inset to Fig.~\ref{Fig6}(b).
The $T$ dependence of $B_c$ is particularly useful, as it allows us to determine the localization length: $\alpha^{-1}$ = $1.4$ and $2.1$ nm at $T =5 \ \rm K$ for samples M90-2 and M90-Pt, respectively.

Using the above-mentioned values for $\alpha^{-1}$, we deduce $N(E_{F}) \approx 10^{18} \rm \ eV^{-1}.cm^{-3}$ for samples M90-2 and M90-Pt. The calculated values of $N(E_{F})$, listed in Table II, are two orders of magnitude lower than typical values found in highly doped PPy films~\cite{Yoon94,Bufon05,Aguilar99,Joo01}. Nevertheless, this low density of states corroborates with the low doping level observed by XPS measurements. Recently, Hulea {\it et al.}~\cite{Hulea05} reported on the correlation between the doping ratio and $N(E_{F})$ in PPy films. They showed that a doping level change from 16\% to 6\% corresponds to a $T_0$ change of about four orders of magnitude. For the lowest doped sample (6\% close to the estimate doping value of PPy in our tri-segmented nanowires), they deduce $T_0 = 5  \times 10^7 \ \rm K$ in agreement with our findings.

\begin{figure}[tbp]
\centering
\includegraphics[width=0.9\columnwidth]{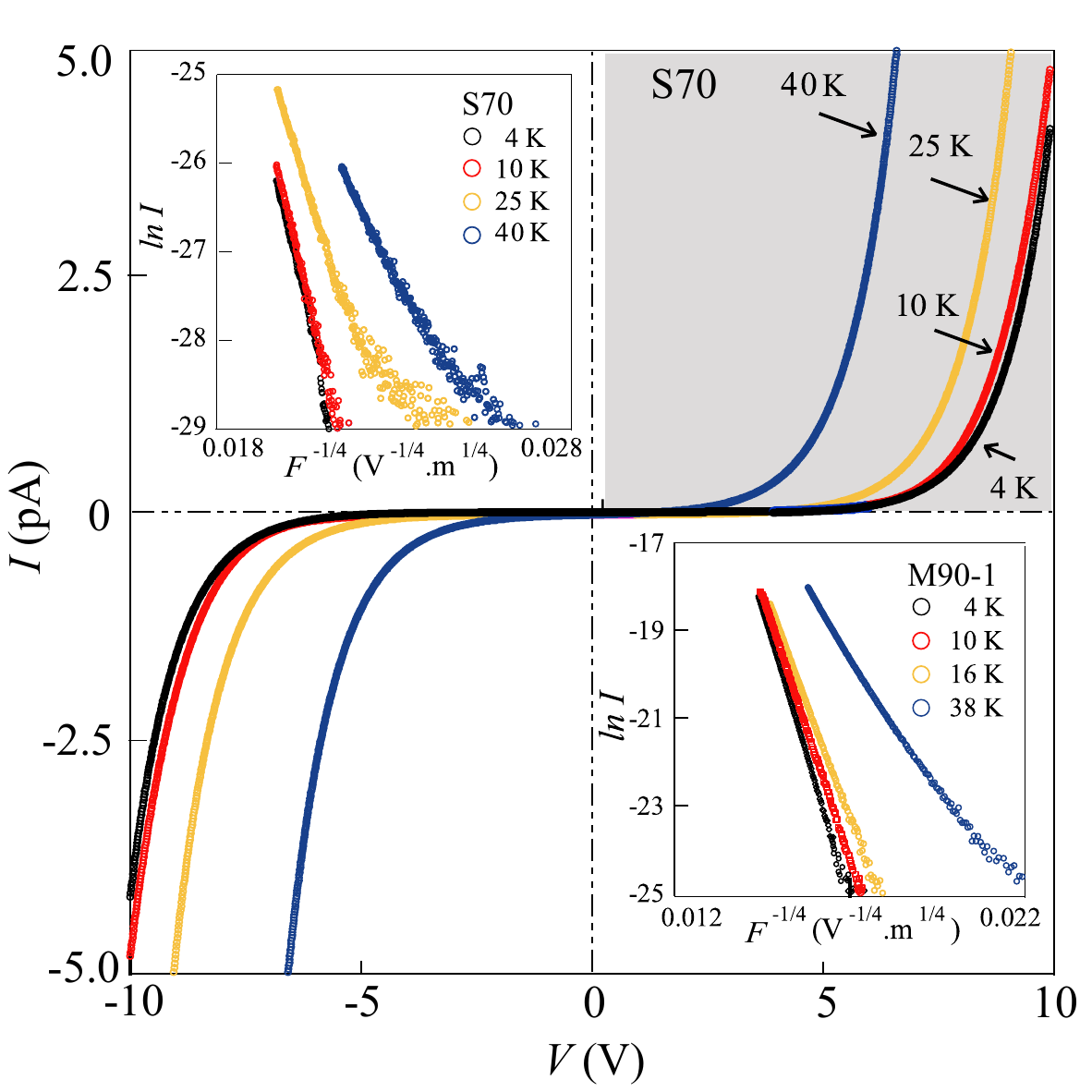}
\caption{[Color online] \IV characteristics at temperatures below 40~K for sample S70. The grey region corresponds to $\beta  \geqslant 1$. The insets display data for samples S70 and M90-1, plotted as $ln\ I$ versus $F^{-1/4}$.}
\label{Fig7}
\end{figure}

\subsection{Electric field-induced hopping}
To further understand the non-ohmic conduction at low $T$, we have analyzed the electric field $F$ dependence of the electrical resistance~\cite{Apsley}.
Indeed, the parameter $\beta = e F \alpha^{-1}/ k_B T$ defines two different hopping transport regimes. For $\beta  \ll 1$, transport is dominated by thermally assisted hopping, while for $\beta  \gg 1$, the electric field-induced hopping mainly contributes to the transport.
Within the 3D electric field-induced hopping regime, the current follows a $ln\ I \propto (F_0/F)^{1/4}$ dependence, with $F_0$ standing for the critical electric field~\cite{Apsley,Shahar90}.
As shown in Fig.~\ref{Fig7}, to reach the $\beta  \gg 1$ regime, data have been taken down to liquid helium temperatures and below. 
In the insets to Fig.~\ref{Fig7}, we plot  $ln\ I$ versus $F^{-1/4}$ for samples S70 and M90-1 at temperatures below 40~K. 
Clearly, a linear behavior is observed at the lowest temperatures, which remains unchanged upon further decrease of $T$.
The parameter $F_0$ could  be extracted from a linear fit to the lowest $T$ data.
The deduced values for samples S70, M90-1, M90-2, and M90-Pt are included in Table~\ref{Tab2}.

The critical electric field $F_0$ could be expressed as  $F_0 = Ck_BT_0/e\alpha^{-1}$, where $C = 8/3$~\cite{Apsley}.
From $F_0$ values we can estimate the localization length $\alpha^{-1}(F_0,T_0)$ for samples S70, M90-1, M90-2 and M90-Pt, as presented in Table~\ref{Tab2}. These values can be compared to those obtained from the magnetoresistance analysis for samples M90-2 and M90-Pt. A reasonable agreement is found, which further substantiates our conclusion that the transport mechanism in our samples is the 3D-VRH.

\subsection{Electronic properties - nanowire structure relationship}

Finally, we discuss our transport results in the light of the structural analysis of hybrid nanowires. 
As function-structure relationships are at the heart of many nanowire synthesis efforts, we address in particular the geometric behavior and the role of the metal-PPy interfaces in our measurements.
In principle, for multi-nanowire samples, one expects a wide distribution of the polymer segment and of the quality of the interfaces.  These disparities between nanowires should result in an averaging effect on measurements and can contribute to mask the intrinsic electronic properties of a single nanowire.
We have shown data for PPy nanowires of two different diameters and two different lengths: 1500 and 300 nm. Even assuming a reasonable 50\% spectral distribution of the grown segments, our data covers  {\em distinct} geometries for the PPy nanowires (see, for instance, samples M70-1 and M90-2). Despite the unavoidable average effect in such multi-wire, vertical geometry samples, parameters extracted from the VRH model are essentially the same to those deduce for planar single nanowires, indicating that we probe the intrinsic behaviour of the PPy. Moreover, we have repeated these measurements in membranes with a pore density of $10^{8} \rm /cm^{2}$ and found similar results.

The carrier injection at  metal-organic interfaces could be modifed by assembling thiol monolayers on the metal electrodes. Hence, in the presence of thiol monolayers adsorbed onto the metallic segments, PPy nanowires could have a different transport signature.
In membranes with nanopores of 70 nm diameter, we have assembled a monolayer of $\rm  SH-(CH)_2-COOH$ onto the first grown Au segment, prior to the electrodeposition of the PPy segment.
Measured \IV characteristics  as a function of $T$ are exemplified in  the lower inset of Fig.~\ref{Fig4}(b) for sample M70-2. They are similar to those measured for tri-segmented nanowires. The only notable difference  is that the zero-bias resistance increases by cca. 3 orders of magnitude as it represents the sum of PPy nanowires' resistance and that of the molecular junctions at the bottom interface. The dominant transport mechanism in molecular junctions is the coherent non-resonant tunneling~\cite{Mbindyo02}. Our data corroborates with the previous results on carboxylate-terminated alkanethiols~\cite{Mbindyo02} and reveal  that both organic sections present in  such multi-segmented nanowires result in clear, unmistakable electrical signatures.

\section{CONCLUSIONS}
Well-shaped tri-segmented Au-PPy-Au and Pt-PPy-Au nanowires presenting different polymer in-wire junction length were prepared by an all-electrochemical process. One major advantage of the exclusively electrochemical process to prepare hybrid nanowires is that the length of each segment as well as the total length of the hybrid nanowires can be easily varied. Characterization by TEM of the polymer junctions unambiguously demonstrates the absence of short-circuits but reveals a strong structural difference between the bottom polymer-onto-metal interface and the upper metal-onto-polymer interface, depending on the choice of materials.  The low $T$ electrical properties of these hybrid nanowires were investigated. At room temperature the \IV characteristics are ohmic.
The  conductivity of about $0.04\ \rm S.cm^{-1}$ obtained for single nanowires is in agreement with the rather low PPy doping ratio (less than 13\%) and is comparable to that of bulk insulating polypyrrole.
Variable temperature measurements of tri-segmented hybrid nanowires confined in the PC template corroborate with single nanowire data. The transport properties in all the samples are consistent with the 3D-VRH model. Our understanding of the non-ohmic 3D hopping transport is consolidated by the low $T$ magnetoresistance data. We have explored nanowires with diameters of 70 and 90 nm, and the electronic properties remained insensitive to the diameter.  For the synthesis of smaller diameter hybrid metal-PPy-metal nanowires, Pt-PPy-Au are the most promising candidates.
We believe that the present technique could be exploited in the fabrication of vertical organic nanowire field-effect transistors incorporated into self-supported flexible nanostructured polymer foils as a template as well as in the design of novel nanowire-based biosensors. 

\begin{acknowledgments}
The work was supported by the FNRS and the NANOMOL project ["Actions de recherches concert\'ees (ARC) - Communaut\'e fran\c{c}aise de Belgique"]. L.G. acknowledges financial support from F.R.I.A. We are indebted to Etienne Ferain and it4ip company for supplying polycarbonate membranes.
\end{acknowledgments}


\begin{thebibliography}{}
\bibitem{Review} For a review see, P.A. Serena and N. Garcma eds., \emph{Nanowires}, (Kluwer, Dordrecht, 1997); Z.L. Wang ed., \emph{Nanowires and Nanobelts}, (Springer, New York, 2005).

\bibitem{Cui01a} Y. Cui {\it et al.}, Science \textbf{293}, 1289 (2001).

\bibitem{Cui01b} Y. Cui {\it et al.}, Science \textbf{291}, 851 (2001).

\bibitem{HJChung05} H.-J. Chung {\it et al.}, Appl. Phys. Lett. \textbf{86}, 213113 (2005).

\bibitem{Husain03} A. Husain {\it et al.}, Appl. Phys. Lett.  \textbf{83}, 1240 (2003).

\bibitem{Koezuka83} H. Koezuka {\it et al.}, J. Appl. Phys. \textbf{54}, 5 (1983).

\bibitem{Nishizava95} M. Nishizava {\it et al.}, Science \textbf{268}, 700 (1995);
C. Schonenberger {\it et al.}, J. Phys. Chem. B \textbf{101}, 5497 (1997).

\bibitem{Mbindyo02} J.K.N. Mbindyo {\it et al.}, J. Am. Chem. Soc. \textbf{124}, 4020 (2002).

\bibitem{Park04} S. Park {\it et al.}, J. Am. Chem. Soc. \textbf{126}, 11772 (2004).

 \bibitem{Long05} Y. Long {\it et al.}, Phys. Rev. B \textbf{71}, 165412 (2005).

\bibitem{Aleshin04} A.N. Aleshin {\it et al.}, Phys. Rev. Lett. \textbf{93}, 196601 (2004);
{\it  ibid.}, Phys. Rev. B \textbf{69}, 214203 (2004).

\bibitem{Park02b} J.G. Park {\it et al.}, Appl. Phys. Lett. \textbf{81}, 4625 (2002).

\bibitem{He03} H.X. He {\it et al.}, Phys. Rev. B \textbf{68}, 045302 (2003).

\bibitem{Saha02} S.K. Saha {\it et al.}, Appl. Phys. Lett. \textbf{81}, 3645 (2002).

\bibitem{Reynes05} O. Reynes and S. Demoustier-Champagne, J. Electrochem. Soc. \textbf{152}, D130 (2005).

\bibitem{Dauginet04} L. Dauginet - De Pra, PhD Thesis, Universit\'{e} catholique de Louvain, Louvain-la-Neuve, 2004.

\bibitem{it4ip} The polycarbonate membranes are available at www.it4ip.be

\bibitem{Duvail} J.L. Duvail {\it et al.}, Synthetic Metals \textbf{131}, 123 (2002);
J.L. Duvail {\it et al.}, J. Phys. Chem. B \textbf{108}, 18552 (2004).

\bibitem{Choy77} C. L. Choy and D. Greig, J. Phys. C: Solid State Phys., \textbf{10},  169 (1977).

\bibitem{Substrates} Silicon wafers [100], p - type, 12 - 25 $\Omega½$.cm, purchased from NEYCO s.a. were used as substrates for the fabrication of planar devices.

\bibitem{Kang93} E.T. Kang {\it et al.}, Adv. Polym. Sci. \textbf{106}, 135 (1993).

\bibitem{Dauginet05} L. Dauginet - De Pra and S. Demoustier-Champagne, Polymer \textbf{46}, 1583 (2005).

\bibitem{fit} Curve fitting has been done using a Gaussian-Lorentzian (85-15\%) linear combination and a linear background.

\bibitem{tri} The PPy content of tri-segmented Au-PPy-Au nanowires, and hence the XPS signal, was too low to obtain a reliable doping ratio.

\bibitem{Sze81} S.M. Sze, {\it Physics of Semiconductor Devices},  (Wiley, New York, 1981), 2nd ed.

\bibitem{Shen06} J.Shen {\it et al.}, Appl. Phys. Lett. \textbf{88}, 253106 (2006).

\bibitem{Mott79} N.F. Mott and E.A. Davis, {Electronic Processes in Non-Crystalline Materials},  (Clarendon Press, Oxford, 1979).

\bibitem{Shklovskii84} B.I. Shklovskii and  A.L. Efros, {\it Electronic Properties of Doped Semiconductors}, (Springer, Berlin, 1984).

\bibitem{MR} We have measured the magnetoresistance of multi-wire samples as the strength of the signal is too weak in single nanowires.

\bibitem{Yoon94} C.O. Yoon, M. Reghu, D. Moses and A.J. Heeger, Phys. Rev. B \textbf{49}, 10851 (1994).

\bibitem{Bufon05} C.C.B. Bufon {\it et al.}, J. Phys. Chem. B \textbf{109}, 19191 (2005).

\bibitem{Aguilar99} J. Aguilar-Hern\'andez {\it et al.}, Phys. Chem. Chem. Phys. \textbf{1}, 1735 (1999).

\bibitem{Joo01} J. Joo {\it et al.}, Synthetic Metals \textbf{117}, 45 (2001).


\bibitem{Hulea05} I. N. Hulea {\it et al.}, Phys. Rev. B \textbf{72}, 054208 (2005).

\bibitem{Shahar90} D. Shahar and Z. Ovadyahu, Phys. Rev. Lett. \textbf{64}, 2293 (1990).
 
\bibitem{Apsley} N. Apsley and H. Hughes, Phil. Mag. \textbf{30}, 963 (1974);
 N. Apsley and H. Hughes, Phil. Mag. \textbf{31}, 1327 (1975).

\end{thebibliography}
\end{document}